\documentclass[11pt,a4paper]{article}

\usepackage{mathpple}
\usepackage{graphicx}
\usepackage{url}
\usepackage{paralist}
\usepackage{textcomp}

\usepackage{geometry}
\begin{document}

\noindent
\textbf{Preprint of:}\\
W. Singer, T. A. Nieminen, N. R. Heckenberg
and H. Rubinsztein-Dunlop\\
``Nanotrapping and the thermodynamics of optical tweezers''\\
in R. Sang and J. Dobson (eds),
\textit{Australian Institute of Physics (AIP)
  17th National Congress 2006: Refereed Papers},
Australian Institute of Physics, 2006 (CD-ROM, unpaginated).

\hrulefill

\begin{center}

{\huge
\textbf{Nanotrapping and the thermodynamics\\ of optical tweezers}}

\vspace{2mm}

\textit{W. Singer, T. A. Nieminen, N. R. Heckenberg
and H. Rubinsztein-Dunlop}\\
School of Physical Sciences, The University of Queensland, Australia

\subsection*{Abstract}

\end{center}

\begin{quote}
Particles that can be trapped in optical tweezers range from tens of
microns down to tens of nanometres in size. Interestingly, this size
range includes large macromolecules. We show experimentally, in
agreement with theoretical expectations, that optical tweezers can be
used to manipulate single molecules of polyethylene oxide suspended in
water. The trapped molecules accumulate without aggregating, so this
provides optical control of the concentration of macromolecules in
solution.
Apart from possible applications such as the micromanipulation of
nanoparticles, nanoassembly, microchemistry, and the study of biological
macromolecules, our results also provide insight into the thermodynamics
of optical tweezers.
\end{quote}

\section*{Introduction---the Limits of Optical Trapping}

While optical tweezers (Ashkin~\textit{et~al.}~1986) are usually used for the
trapping and manipulation of particles of approximately 1--10\,{\textmu}m
in size, both larger and smaller particles can be trapped. Since an
upper limit to the force that can be exerted by the beam is imposed by
the finite momentum flux of the beam, while weight, viscous drag, and the
likelihood of adhesion all continue to increase with increasing size,
optical tweezers will become unable to perform useful manipulation for
sufficiently large particles.

For small particles, the gradient force becomes proportional to
the volume, while the scattering forces become proportional to
the volume squared (Svoboda and Block~1994; Harada and Asakura~1996).
Therefore, a wider range of particles become trappable at small
sizes, such as reflective particles that are pushed out of the trap
at larger sizes. Absorptive particles remain untrappable at small
sizes since the absorption force is proportional to the volume.

The lower limit is provided by Brownian motion------the Brownian
motion force is $12\pi r \eta k_B T$ where
$\eta$ is the viscosity of the fluid and $k_B$ is
Boltzmann's constant. Since the viscosity of the fluid will
hinder escape, it is best to compare the trapping
potential---the energy required to remove the particle from
the trap---with the thermal energy $k_B T$.

Since the gradient force acting on a small sphere is
given by (Harada and Asakura~1996)
\begin{equation}
\mathbf{F}_\mathrm{grad} = \frac{2\pi n_\mathrm{med} a^3}{c}
\left( \frac{m^2-1}{m^2+2} \right) \nabla I
\label{gradient_force}
\end{equation}
where $n_\mathrm{med}$ is the refractive index of the
surrounding medium, $m = n_\mathrm{particle}/n_\mathrm{med}$
is the relative refractive index, $c$ is the speed of light
in free space, and $I$ is the irradiance $I$, the corresponding
trapping potential is equal to
\begin{equation}
U = \frac{\pi n_\mathrm{med} a^3}{c}
\left( \frac{m^2-1}{m^2+2} \right) I.
\label{trapping_potential}
\end{equation}
This suggests
that it should be possible to trap polymer microspheres as
small as 30\,nm in radius with a power as low as
100\,mW available at the focus. As this size range includes
single polymer macromolecules, we carried out experiments
on the optical trapping of single polyethylene oxide (PEO)
molecules of differing molecular weights (Singer~\textit{et~al.}~2006).

\section*{Experiment and Experimental Results}

A typical inverted optical tweezers apparatus, with
a $100\times$ $\mathrm{NA}=1.25$ objective, was 
used to focus the beam from a 1064\,nm ytterbium
fiber laser to form the trap. Up to 700\,mW was
available at the focus.
A sample consisting of a suspension of PEO molecules
was place in the trap, and the concentration of PEO
molecules in the trap was monitored by measuring the
backscattered light from a low-power (2\,mW) He--Ne
laser focused onto the same position. The properties
of the PEO molecules are given in tables 1 and 2.
Notably, a comparison of measured radii of gyration
(Devanand and Selser~1991) with the stretched chain
lengths (Cooper \textit{et al.} 1991) shows
that the PEO molecules can be approximated as
spheres, and that the above theoretical considerations
should apply.

\textbf{Table 1. Concentrations of PEO solutions}
\begin{center}
\begin{tabular}{cccc}
molecular weight (kDa) & number density (($\mu$m)$^{-3}$)
& mass fraction & concentration:overlap\\
\hline
100 & 313 & 0.01\%  & 0.02 \\
300 & 281 & 0.027\% & 0.07 \\
900 & 268 & 0.077\% & 0.48 \\
\end{tabular}
\end{center}

\textbf{Table 2. Radii of gyration and stretched chain
lengths of PEO molecules}
\begin{center}
\begin{tabular}{ccc}
molecular weight (kDa) & radius of gyration (nm)
& stretched chain length ($\mu$m) \\
& (Devanand and Selser 1991) & (Cooper \textit{et al.} 1991) \\
\hline
100 & 17.6 & $-$ \\
300 & 33.5 & 2.4 \\
900 & 63.6 & 7.2 \\
\end{tabular}
\end{center}

The two larger sizes of PEO molecules (100\,kDa and 300\,kDa)
could be trapped. When the molecules were trapped the concentration
of PEO molecules was observed to increase over several minutes.
A typical experimental curve is shown below in figure 1.

\centerline{\includegraphics[width=0.6\columnwidth]{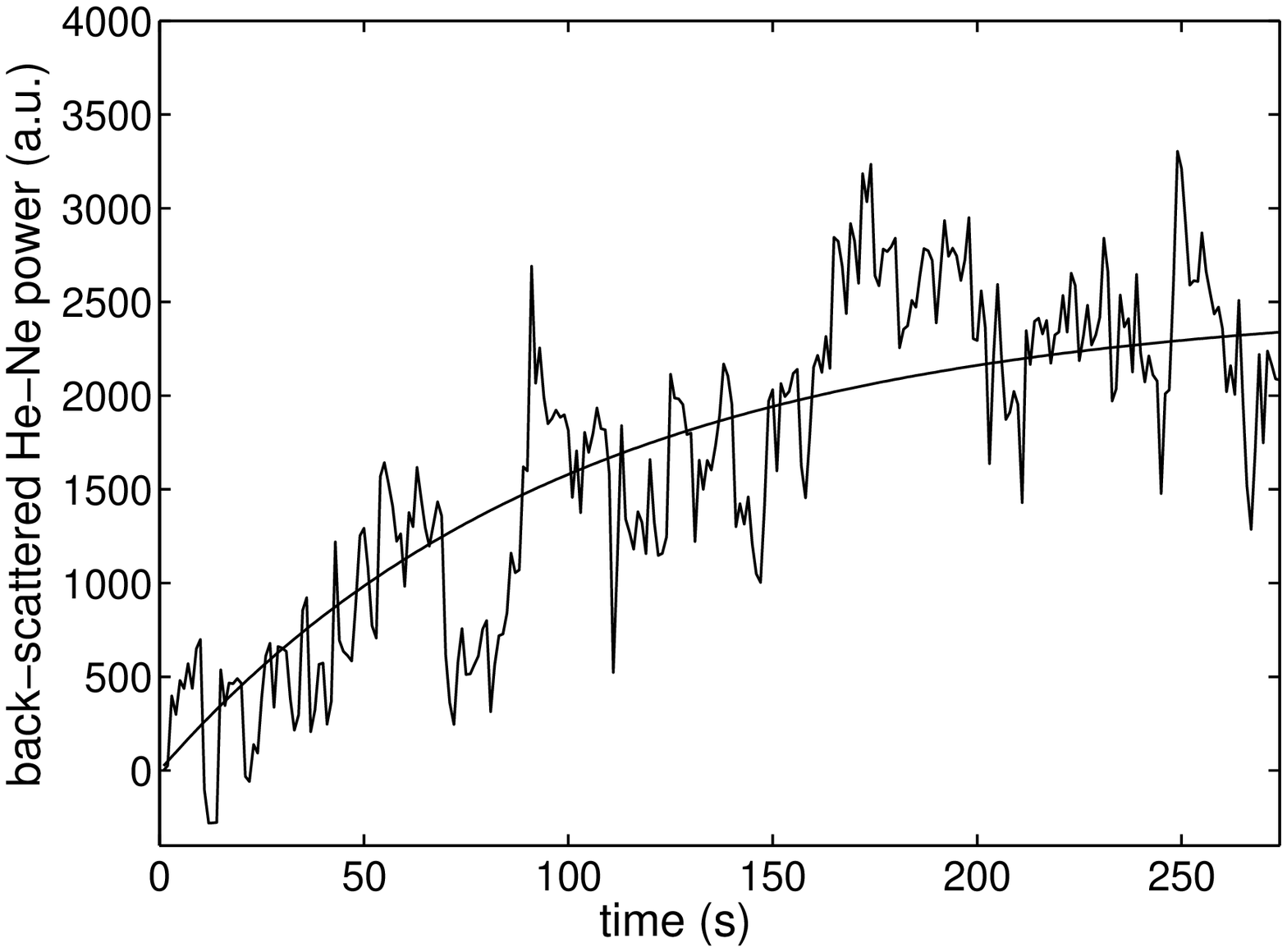}}

\textbf{Figure 1. Increase in concentration over time}

The increase in concentration was repeatable, reversible,
and increased with increasing power (Singer~\textit{et~al.}~2006).
We estimate that approximately
1.5\,W would be required to trap the 100\,kDa PEO molecules.

\section*{Thermodynamics of Optical Tweezers}

The increase in concentration takes several minutes
because PEO molecules must diffuse into the region of the trap.
As the trap is initially turned on, the molecules around the edge of
the trap are rapidly pulled into the middle of the trap. This leaves
a region of reduced concentration into which molecules outside the trap
will diffuse into.

In the dilute limit, the probability of
finding a particle in a particular position will
follow a Boltzmann distribution. Since the ambient
energy is $k_B T$, and a particle within the
trap has its energy reduced by the trapping
potential, the probability per unit volume of finding a particle
within the trap is greater, and the equilibrium concentration
will be
\begin{equation}
C(\mathbf{r}) = C_0 \exp( -U(\mathbf{r})/k_B T ),
\end{equation}
where $C_0$ is the ambient concentration and $U$ is
the trapping potential. As soon as the laser power is
sufficient for the trapping potential to exceed
$k_B T$, the concentration increases very rapidly with
increasing power. This results in the appearance of
sudden switching-on at a threshold power---also
seen in other systems, such as, for example, diodes in circuits,
which have similar exponential behaviour.

At equilibrium, there are two opposing forces acting on the molecules:
the optical gradient force, and a force due to the gradient of
the partial pressure of the molecules. At equilibrium,
\begin{equation}
F_\mathrm{grad}C = RT\nabla C
\end{equation}
where $R$ is the universal gas constant (Einstein~1905; Einstein~1956).
Before equilibrium is reached,
the imbalance between these two forces will drive diffusion of the
PEO molecules. Assuming spherical symmetry of the trap, this
reduces to a one-dimensional differential equation and
can be solved using a finite-difference time-domain method.
As the times scales involved---the time over which the optical
force moves the molecules, and the time taken to diffuse into
the region of the trap---are very different, the simulation
requires a very large number of time steps to show the
behaviour seen in the experiments. The simulation was carried
out with a weak optical force to reduce the discrepancy between
the two time scales. Results are shown in figures 2 and 3.

\includegraphics[width=0.32\columnwidth]{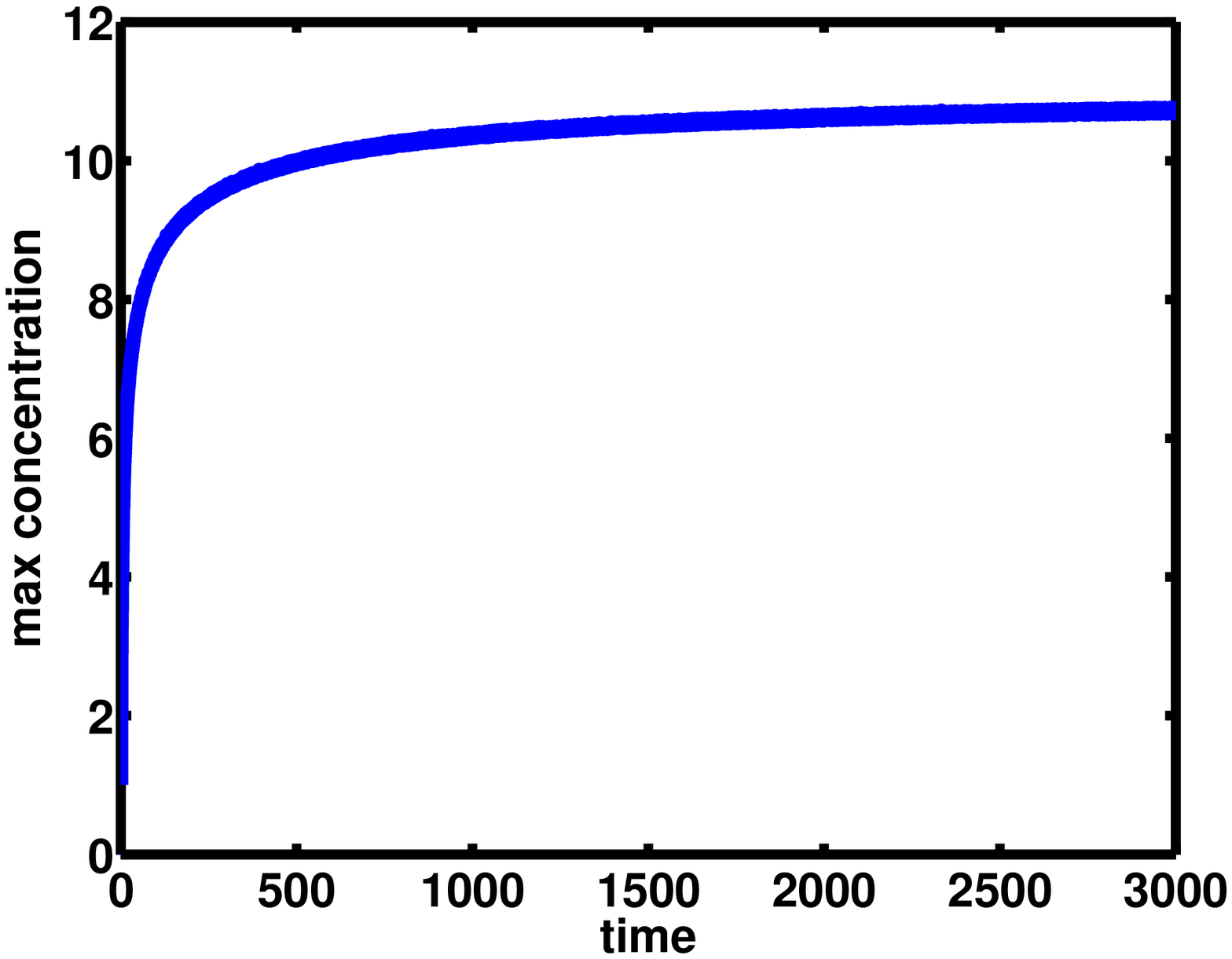}
\hfill
\includegraphics[width=0.32\columnwidth]{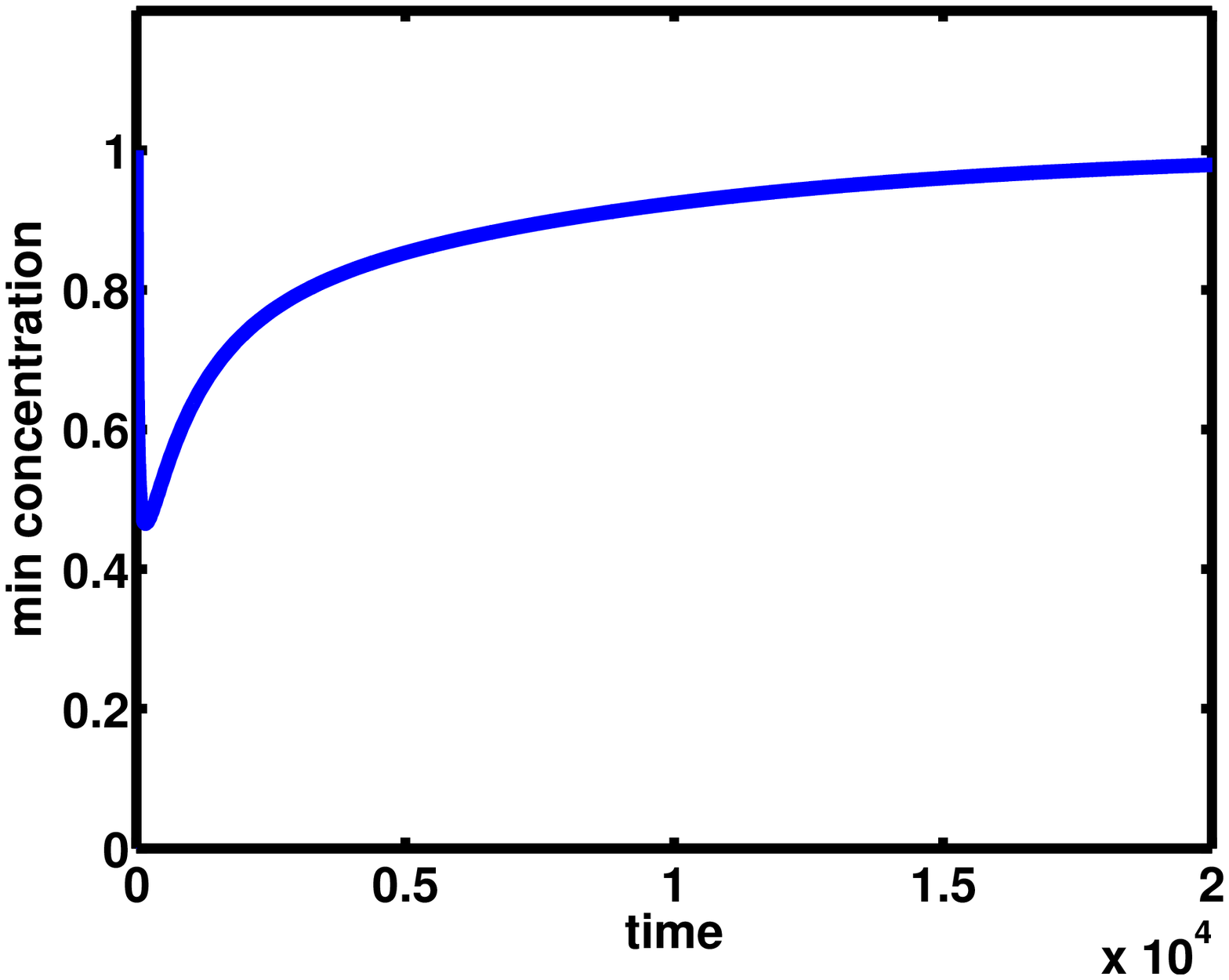}
\hfill
\includegraphics[width=0.32\columnwidth]{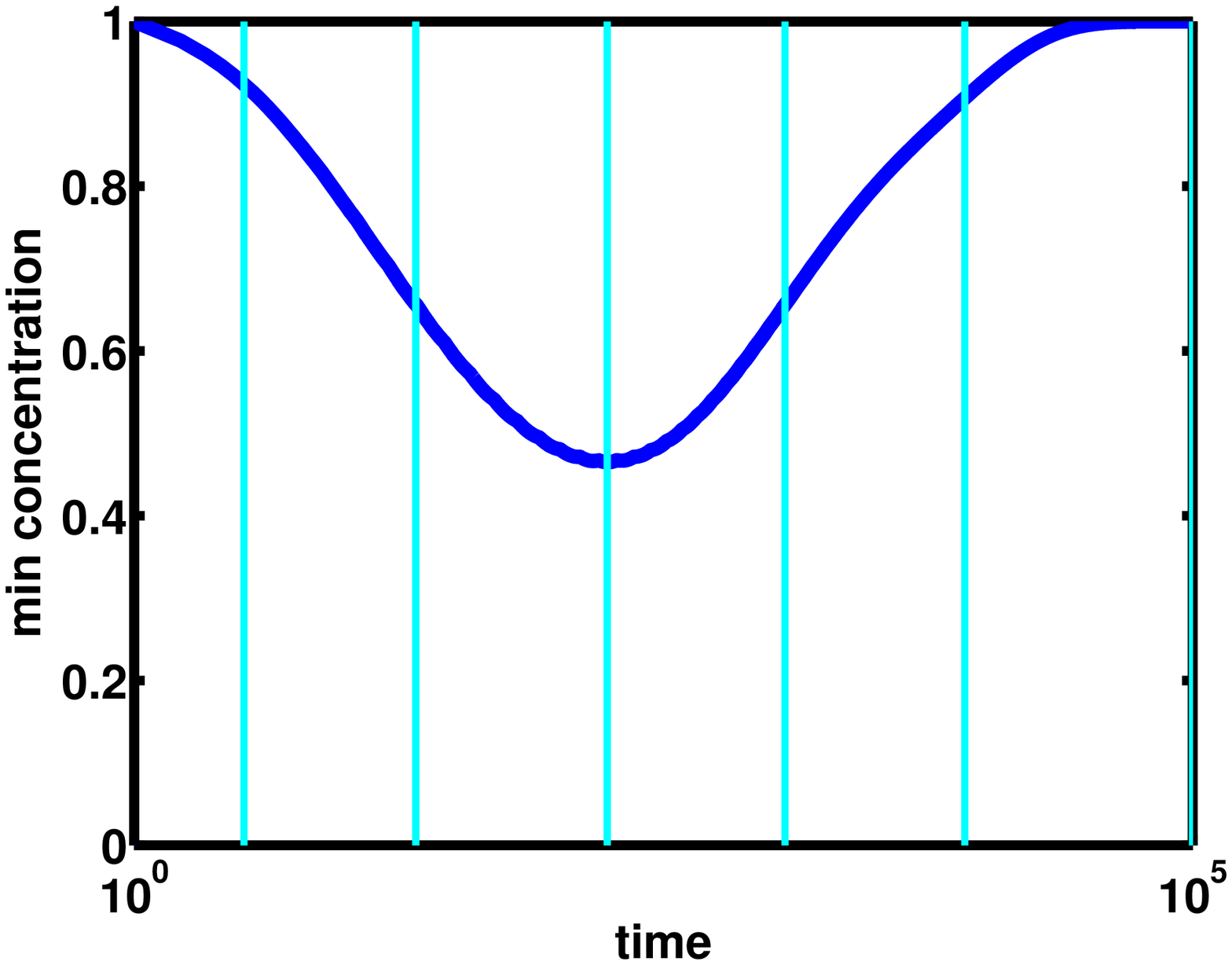}\\
(a) \hfill (b) \hfill (c) \hfill ~

\textbf{Figure 2. Change in concentration over time.} (a) shows the
growth of the maximum concentration over time, (b) shows the
change in the minimum concentration over time. (c) shows the
times at which the concentration profiles shown in figure 3
occur. 

\includegraphics[width=0.32\columnwidth]{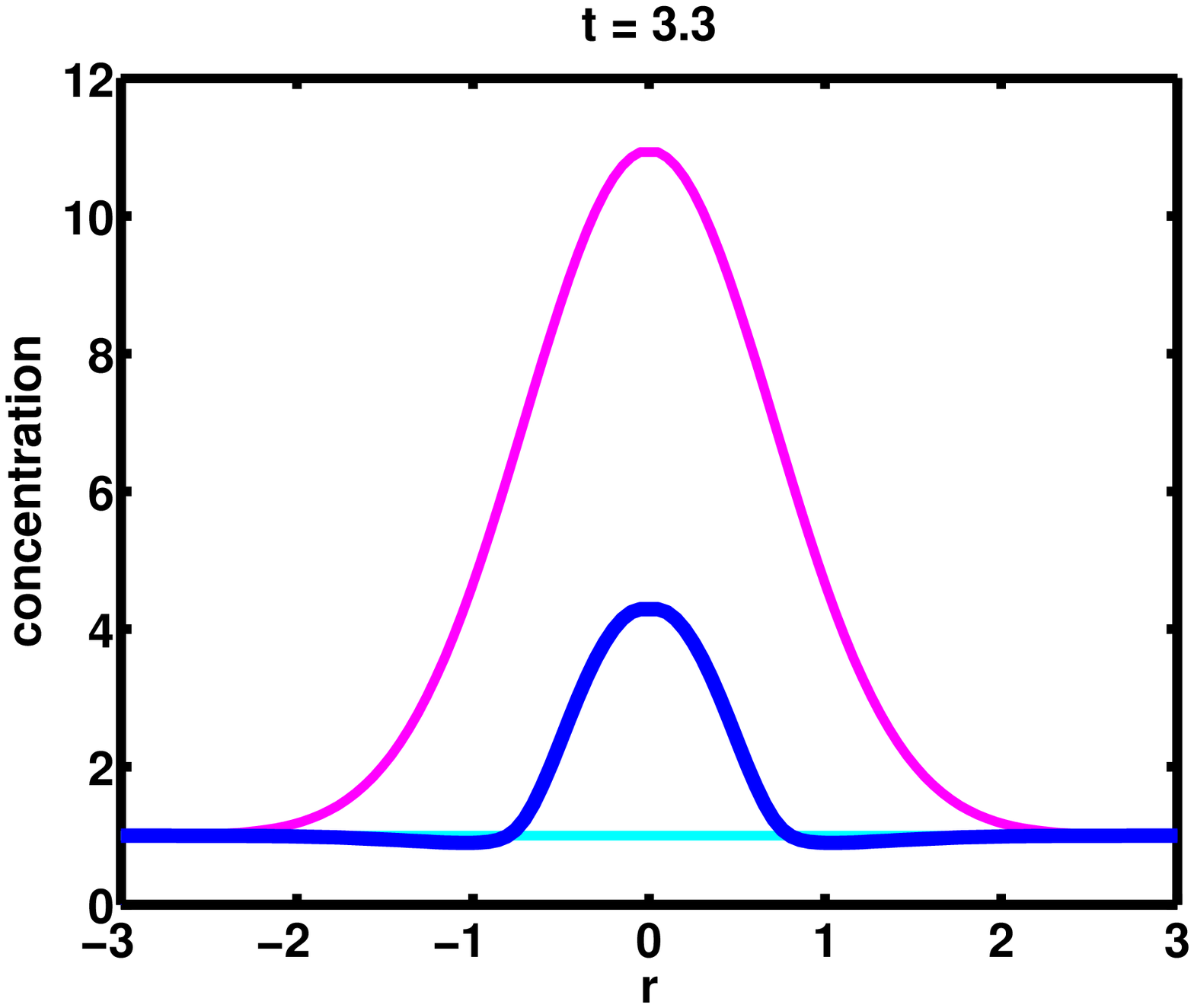}
\hfill
\includegraphics[width=0.32\columnwidth]{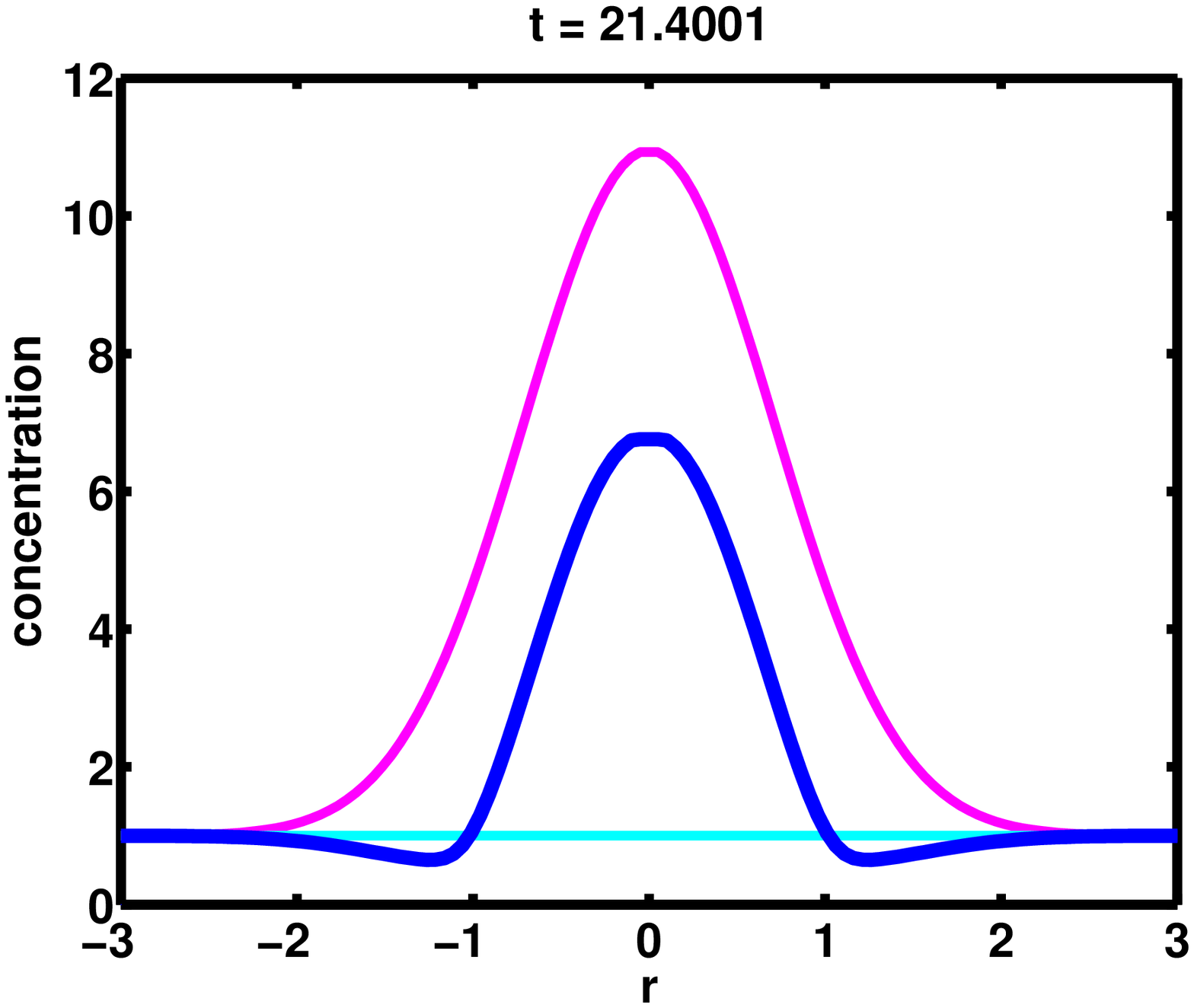}
\hfill
\includegraphics[width=0.32\columnwidth]{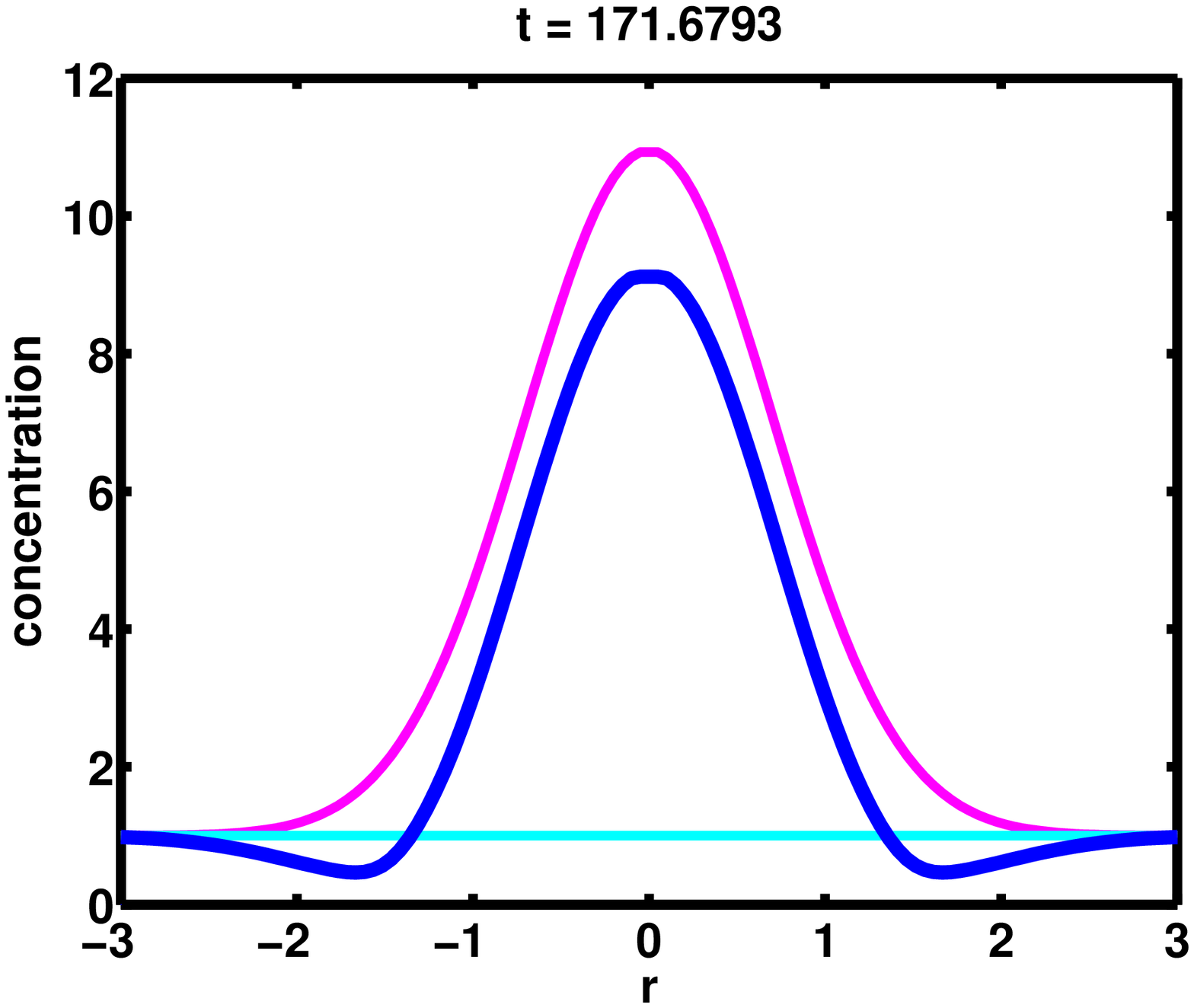}\\
(a) \hfill (b) \hfill (c) \hfill ~\\
\includegraphics[width=0.32\columnwidth]{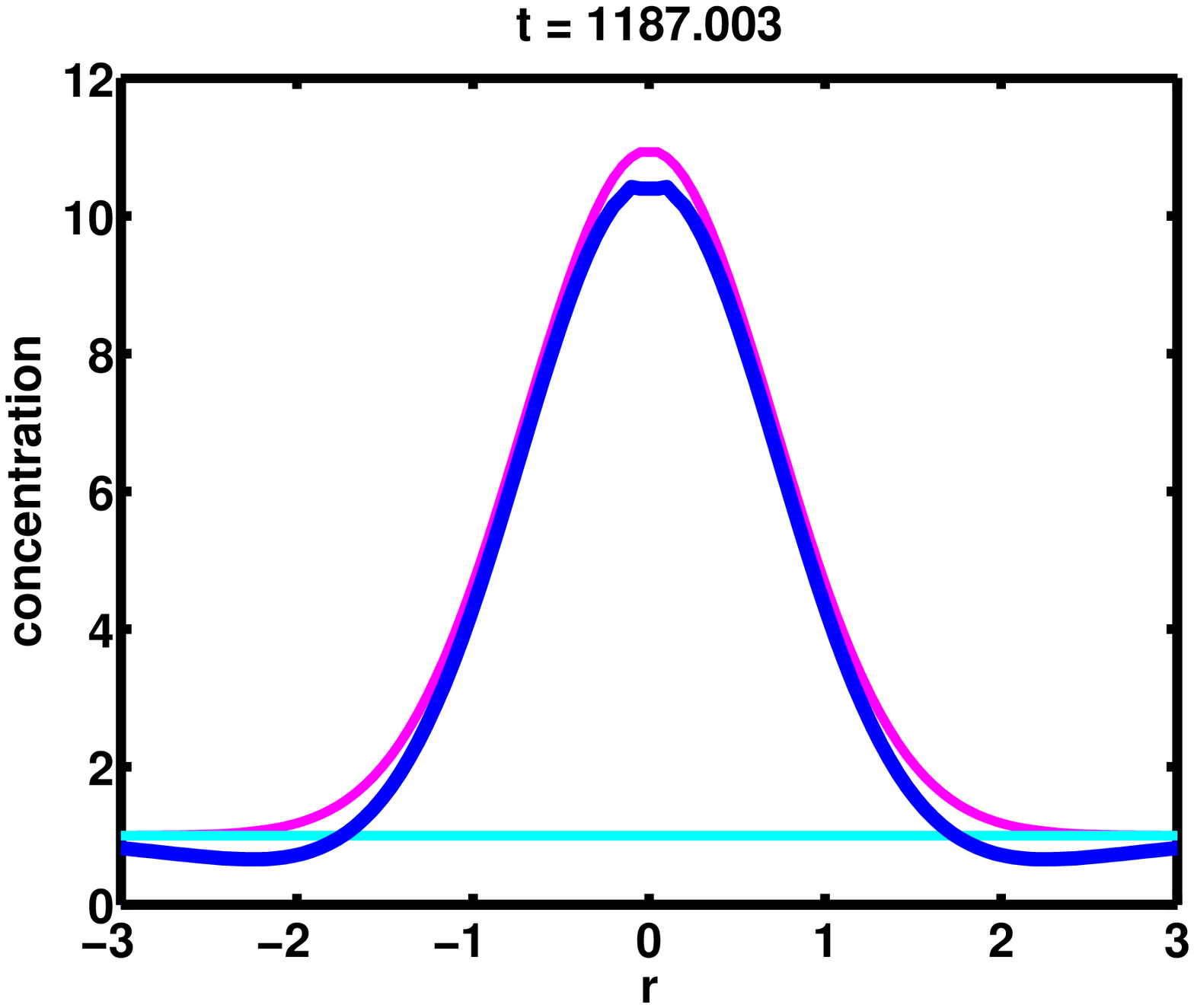}
\hfill
\includegraphics[width=0.32\columnwidth]{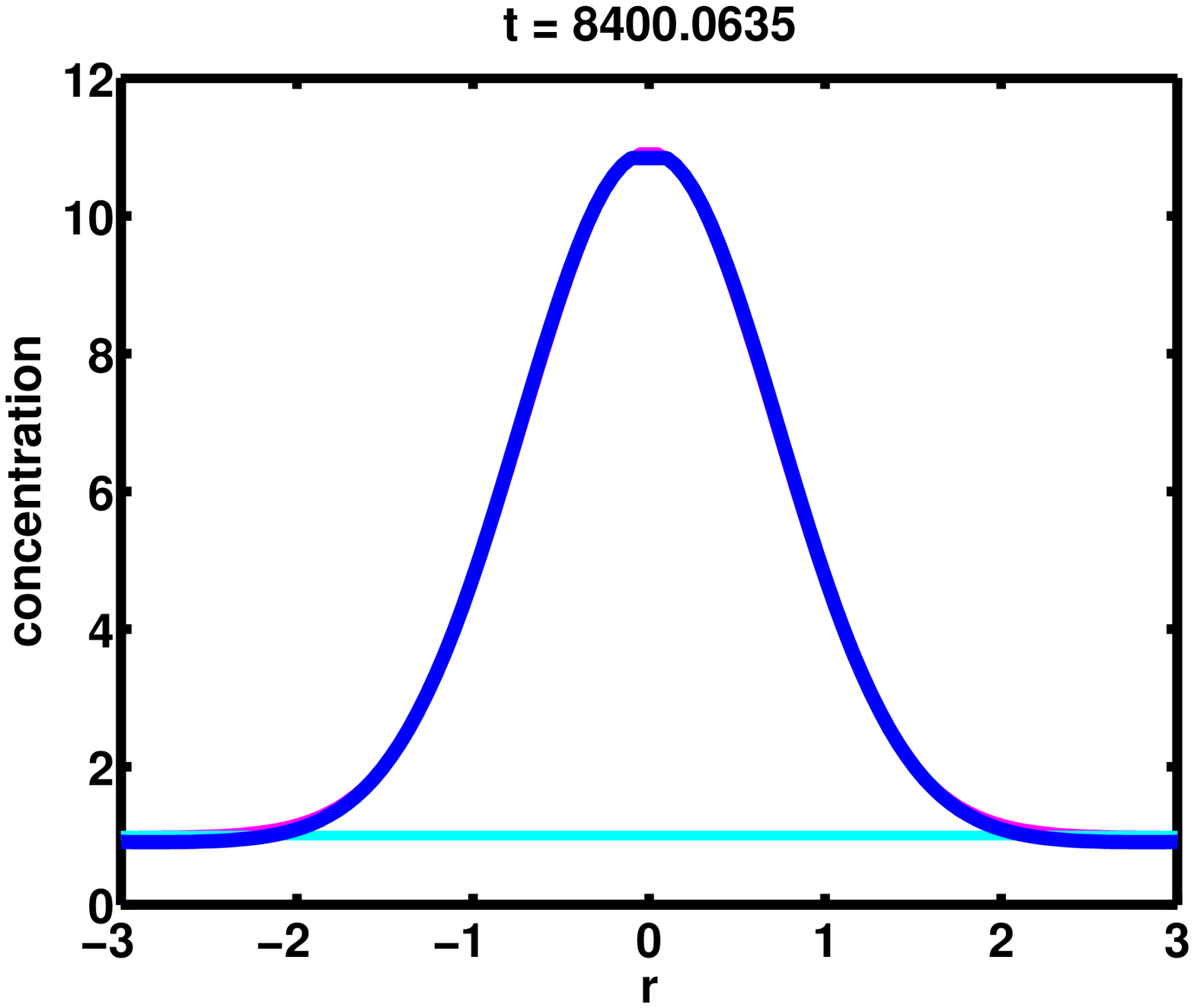}
\hfill
\includegraphics[width=0.32\columnwidth]{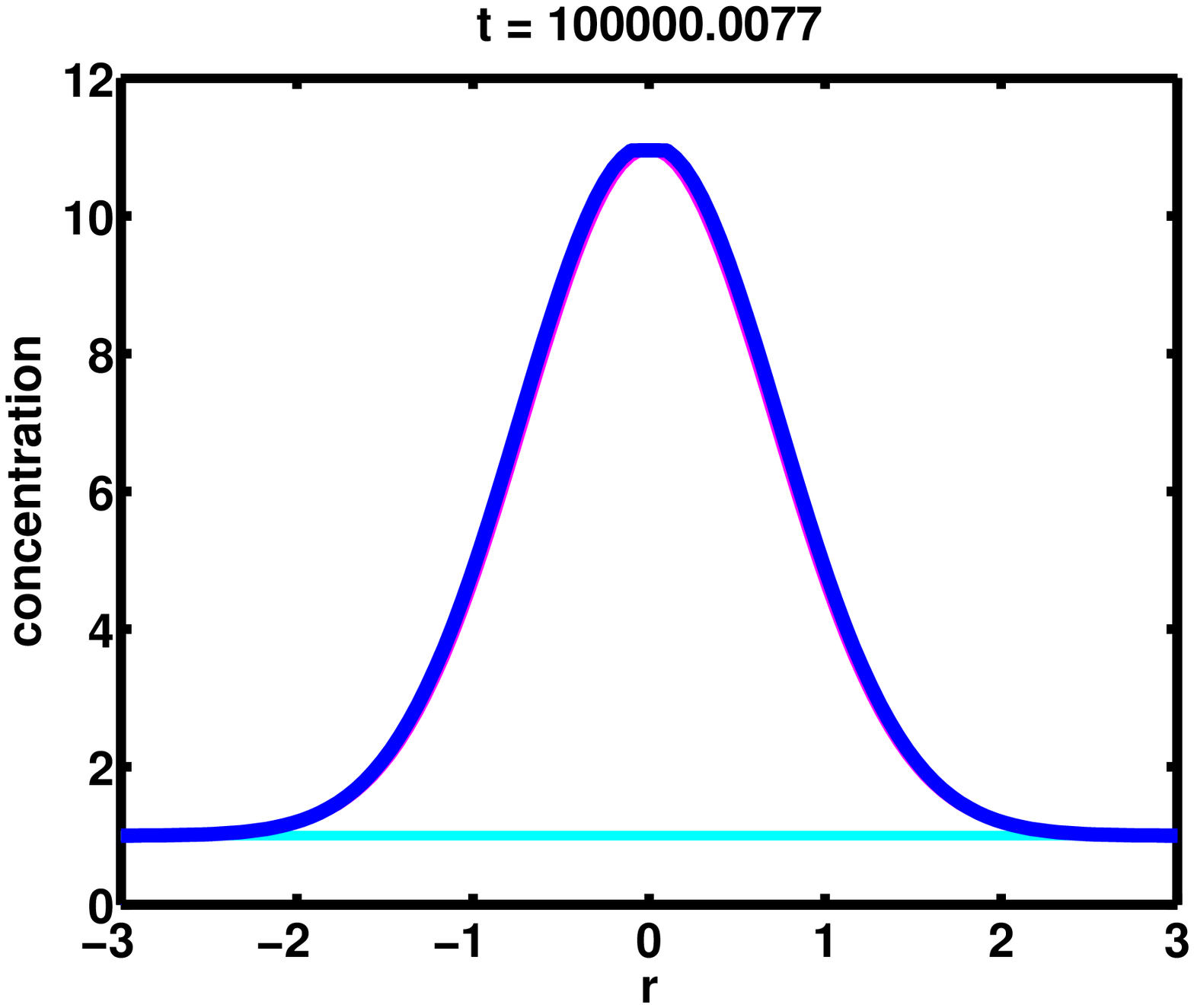}\\
(d) \hfill (e) \hfill (f) \hfill ~

\textbf{Figure 3. Concentration over the trap at different times}

In figure 2(a), we can see the growth in concentration over time.
This is similar to the experimentally observed behaviour in figure~1.
The depletion of concentration in the outer region of the trap is
shown in 2(b). Although we could not directly observe this depletion,
it clearly existed in the experiment, since otherwise PEO molecules
would not have diffused towards the trap. The rapidity of the
initial depletion compared with the slow increase back to the
ambient concentration highlights the difference in the time scales
of the two processes. In order to see how the concentration
changes in more detail, the concentration was plotted as a function
of position for selected times, shown by the light blue lines in
figure 2(c). (Figure 2(c) is otherwise the same as figure 2(b),
except with a logarithmic scale on the horizontal axis.) The
plots of the concentration versus position are shown in figure 3.
The dark blue line shows the instantaneous concentration, with the
light blue line indicating the initial concentration. The magenta
line shows the profile of the trapping beam, which is assumed to be
Gaussian. The initial depletion due to the optical trap concentrating
the immediately available molecules in the centre of the
trap can be clearly seen, followed by a slow approach to equilibrium.

Interestingly, the edge of the optical trap functions as a one-way
barrier for the molecules---they can diffuse into the trap, but
until the partial pressure of the molecules already in the trap is
sufficiently high, they cannot diffuse out. The surface area
of this ``event horizon'' through which the molecules fall
increases with increasing power, which is why the rate of increase
of concentration was observed to increase in the experiments
(Singer~\textit{et~al.}~2006).

\section*{Conclusion}

We have shown that optical tweezers can be used to
control the concentration of macromolecules
in solution. These results illuminate the thermodynamics
of optical tweezers, and provide a striking example of
diffusion under an applied force. This suggests that
optical tweezers may be a useful tool to study the equilibrium and
non-equilibrium thermodynamics of polymer or
other molecules in solution.

\section*{Refererences}

Ashkin, A., Dziedzic, J.M., Bjorkholm, J.E. and Chu, S. (1986).
Observation of a single-beam gradient force optical
trap for dielectric particles. \textit{Optics Letters} \textbf{11}, 288-90.\\
Cooper, E.C., Johnson, P. and Donald, A.M. (1991).
Probe diffusion in polymer-solutions in the dilute semidilute
crossover regime. 1. {P}oly(ethylene oxide).
{\em Polymer} {\bf 32}, 2815-22.\\
Devanand, K. and Selser, J.C. (1991)
Asymptotic-behavior and long-range interactions in aqueous-solutions of
poly(ethylene oxide). {\em Macromolecules} {\bf 24}, 5943-7.\\
Einstein, A. (1905). Uber die von der molekular
kinetischen Theorie der Warme geforderte Bewegung von in ruhenden
Fl\"{u}ssigkeiten suspendierten Teilchen.
\textit{Annalen der Physik} \textbf{17}, 549-60. Reprinted in
\textit{Annalen der Physik} \textbf{14}, Supplement, 182-93 (2005).\\
Einstein, A. (1956) Investigations on the theory of the Brownian movement,
Dover, New York.\\
Harada, Y. and Asakura, T. (1996). Radiation forces on a dielectric sphere
in the {R}ayleigh scattering regime. {\em Optics Communications} {\bf 124},
529-41.\\
Singer, W., Nieminen, T.A., Heckenberg, N.R. and Rubinsztein-Dunlop, H.
(2006). Optical micromanipulation of synthetic macromolecules.
To appear in \textit{Proceedings of SPIE} \textbf{6326}.\\
Svoboda, K. and Block, S.M. (1994). Optical trapping of metallic {R}ayleigh
particles. {\em Optics Letters} {\bf 19}, 930-2.

\end{document}